# Observation and Control of Spontaneous Magnon Emission from Spin Ensembles in 2D Hexagonal Boron Nitride

Ling-Jie Zhou[1], Jayakrishnan M. P. Nair[2], Senlei Li[1], Thomas Poirier[3], Yiran Zhao[1], Zelong Xiong[1], Sumedh Rathi[1], Jingcheng Zhou[1], Zhigang Jiang[1], Hailong Wang[1], James H. Edgar[3], Benedetta Flebus[2,*] and Chunhui Rita Du[1,*]

[1]School of Physics, Georgia Institute of Technology, Atlanta, Georgia 30332, USA
[2]Department of Physics, Boston College, Chestnut Hill, Massachusetts 02467, USA
[3]Tim Taylor Department of Chemical Engineering, Kansas State University, Manhattan, Kansas 66506, USA

[*]Corresponding authors: benedetta.flebus@bc.edu and cdu71@gatech.edu

**Abstract**: Hybrid systems consisting of color centers and magnetic materials provide an appealing solid-state platform for advancing the burgeoning quantum technological revolution. Exploring novel coupling mechanisms between optically active spin defects and quantum degrees of freedom is directly relevant in this context. Here, we report observation and control of spontaneous magnon emission from boron-vacancy centers in 2D hexagonal boron nitride (hBN) to a proximal ferrimagnet in the near-zero temperature limit. The spontaneous magnon emission process is overshadowed by thermal magnon effect as temperature increases, reflecting the crossover from a spontaneous-emission-dominated magnon reservoir to a thermally occupied spin bath where absorption and stimulated processes restore balance. By increasing the spin defect density, we further present that spontaneous magnon emission into a common spin bath could help establish magnon-mediated correlations in dense hBN spin ensembles. Our results are quantitatively captured by theoretical modeling, bringing insights into understanding qubit-magnon coupling, correlated spin dynamics, and many-body physics of color centers in the quantum regime.

**Teaser**: Spontaneous magnon emission-induced asymmetric relaxation of spin defects in the near-zero temperature limit.

## Introduction

Harnessing dipole interaction between color centers and quantum degrees of freedom opens a promising route for implementing cutting-edge technological innovations in quantum metrology, quantum optics, and quantum information science (*1-8*). Exploring advanced material and qubit platforms with improved solid-state scalability, versatile dipole coupling, and novel coupling channels serves as an important step for fulfilling this goal. Optically active spin defects contained in layered van der Waals (vdW) crystals are naturally relevant in this context (*9-12*). Taking advantage of excellent two-dimensional (2D) vdW compatibility and on-chip integrability, this new class of spin defects can be arranged in nanoscale proximity of target materials/devices in an accessible and controllable way, showing great promise for revolutionizing the performance on spatial resolution, field sensitivity, and measurement modality of current quantum sensing technologies (*9-11, 13-22*). To date, experimental detection of magnons, spins, and electric charges in various hard and soft material systems has been demonstrated by hBN-based quantum microscopy (*12, 13, 15, 17, 18, 23-26*).

Despite the remarkable progress made thus far, the ongoing research on hBN spin defects is mostly limited to the conventional $^4$He cryogenic temperatures (*16, 23, 27, 28*), which provides only a subset of information on their intriguing quantum mechanical properties. The underlying spin relaxation mechanism, spin-phonon interaction, spin-spin correlations of hBN color centers and their dipole coupling with other quantum degrees of freedom in sub-Kelvin regime where thermally induced decoherence and energy dissipation are effectively suppressed await for exploration. This knowledge is crucial for designing hybrid systems with desirable quantum spin properties for realizing transformative quantum revolution (*29-31*).

Here, we report dipole coupling observed between boron-vacancy $V_B^-$ centers in 2D hBN and magnons, bosonic quasiparticles carrying spin angular momentum, at millikelvin temperatures. We show that the intrinsic $V_B^-$ spin relaxation rate dramatically decreases by orders of magnitude when the temperature is lowered from ~5 K to ~60 mK. Enjoying the extended spin relaxation times, for the first time, we present experimental evidence of $V_B^-$ spin transition-driven spontaneous magnon emission into a proximal spin bath in the near zero-temperature limit, where thermally induced quantum decoherence and energy dissipation are virtually frozen (*2, 4, 5*). The spontaneous magnon emission process starts to be overshadowed by the thermal magnon effect as temperature increases, reflecting the crossover from a spontaneous emission-dominated, effectively cold magnon reservoir to a thermally occupied bath where absorption and stimulated processes restore balance to erase the quantum signatures. By increasing the spin defect density, we further show that spontaneous magnon emission into a common spin bath from hBN spin ensembles could establish correlations between them, resulting in enhanced magnon-induced $V_B^-$ spin relaxation rate. Our results are quantitatively captured by theoretical models, bringing insights into developing next-generation hybrid quantum magnonic systems composed of solid-state spin defects and magnetic materials (*2, 32*).

## Results

### Spin relaxation of hBN defects in the near zero-temperature limit

We first briefly review the pertinent properties of hBN spin defects used in the current study. Figure 1A shows the atomic structure of boron-vacancy centers formed in the 2D hBN crystal lattice with alternating boron and nitrogen atoms (see Section S1 for details) (*9, 10, 13, 14, 19, 21, 22, 27*). Each $V_B^-$ spin defect is surrounded by three nearest-neighboring nitrogen atoms. The negatively charged $V_B^-$ center possesses an $S = 1$ electron spin with an intrinsic quantization axis along the out-of-plane direction (*9*). Figure 1B plots the energy level diagram of a $V_B^-$ defect. A $V_B^-$ center excited to the $m_s = \pm 1$ spin states is more likely to relax through a non-radiative pathway back to the $m_s = 0$ ground state and emits reduced photoluminescence (PL) (*20, 27, 33*). By measuring PL as a function of the delay time $t$ and fitting the data with an appropriate theoretical model (*34, 35*), occupation probabilities of a $V_B^-$ center can be quantitatively obtained, allowing for extraction of the spin relaxation rate (*16, 23*). Conventional quantum spin relaxometry is typically performed at temperatures high enough to ensure a substantial thermal magnon population in the spin bath, thereby enhancing magnetic noise fluctuations and the corresponding sensor–magnon relaxation signals (*1, 18, 23, 35-38*). However, this regime is not optimal for hybrid quantum magnonic platforms building on spin defects coupled to magnetic materials. The operation of such systems generally favors ultralow temperatures, where the thermal energy scale remains below the spin-defect transition frequency, thereby minimizing decoherence and suppressing unwanted dissipation (*2, 4, 5*).

The current study takes advantage of a millikelvin quantum sensing platform (see Section S2 for details) to investigate spin relaxation dynamics of $V_B^-$ spin ensembles in the quantum regime, i.e., $T \rightarrow 0$, as illustrated in Fig. 1C. For the intrinsic case, $V_B^-$ spin relaxation is dominated by phonon processes in hBN lattices (*13, 16*), presumably showing a slow PL decay rate with an extended spin lifetime. Our spin relaxometry measurements corroborate this point. Figure 1D presents temperature dependence of intrinsic spin relaxation rate $\Gamma_{\mathrm{int}}$ measured on $V_B^-$ spin ensembles with an estimated defect density $\rho_1 = 1.7 \times 10^{17}$ cm$^{-3}$, $\rho_2 = 3.1 \times 10^{17}$ cm$^{-3}$, $\rho_3 = 6.8 \times 10^{17}$ cm$^{-3}$, and $\rho_4 = 1.7 \times 10^{18}$ cm$^{-3}$ contained in hBN nanoflakes. $\Gamma_{\mathrm{int}}$ shows a dramatic decrease when temperature is lowered from 5 K to 60 mK. The temperature dependent $\Gamma_{\mathrm{int}}$ can be fitted to a physically motivated model, suggesting a combination of one-phonon and two-phonon scattering-driven $V_B^-$ spin relaxation (see Section S3 for details). It is worth noting that $V_B^-$ centers exhibit intrinsic spin relaxation rates comparable to that of shallowly implanted nitrogen-vacancy centers in the sub-Kelvin temperature regime (see Section S3 for details). When $V_B^-$ spin defects are coupled with a magnetic material in the zero-temperature limit, there are essentially no thermal magnons populated at the $V_B^-$ spin energy in the magnon bath. In this limit, thermal magnon absorption and stimulated processes are effectively frozen out, as the relevant modes of the spin bath are nearly unoccupied. Instead, in direct analogy with vacuum-induced spontaneous emission in quantum optics, an excited $V_B^-$ center relaxes by emitting a magnon into the quantum vacuum of the magnetic reservoir (Fig. 1C) (*39*). This reveals a spin-defect–magnon coupling channel—driven purely by quantum fluctuations of the bath—that is expected to dominate in the low-temperature regime but has remained experimentally unexplored.

## Observation of spontaneous magnon emission from $V_B^-$ centers

We now present experimental observation of spontaneous magnon emission from $V_B^-$ spin defects in the near zero-temperature limit. Figure 2A shows the schematic of our measurement platform where an exfoliated hBN nanoflake containing $V_B^-$ ensembles is transferred onto an Au microwave stripline patterned on a 110-nm-thick $Y_3Fe_5O_{12}$ (YIG) film. Here, ferrimagnetic insulator YIG serves as a solid-state magnon bath proximal to $V_B^-$ spin defects (*40, 41*). An optical image in Fig. 2B provides an overview of a prepared hBN/YIG device for spin relaxometry measurements. Thickness of the hBN flake is ~75 nm (see Section S1 for details) and microwave signals delivered into the on-chip Au stripline are utilized to provide local control of $V_B^-$ spin states (*23*). Note that the $V_B^-$ spin defect density $\rho$ is estimated to be $3.1 \times 10^{17}$ cm$^{-3}$ and the thickness of the Au stripline $t_{\mathrm{Au}}$ is 110 nm (see Method section for details). Next, we report systematic spin relaxometry measurements on the hBN/YIG device in the sub-Kelvin regime. Figure 2C presents temperature dependent $V_B^-$ spin relaxation rate Γ measured from 58 mK to 300 mK (see Section S4 for details). It is clear that the magnon-$V_B^-$ coupling significantly enhances the $V_B^-$ spin relaxation rate Γ compared with the intrinsic one.

We have developed a detailed theoretical model to capture the transition from high-temperature thermal magnon effect to low-temperature spontaneous magnon-emission-induced $V_B^-$ spin relaxation. Our discussion starts with a generic physical picture of dipole coupling between $V_B^-$ centers and a YIG magnon bath. At thermal equilibrium, distribution of YIG magnons statistically follows the Bose-Einstein function: $n_B = (e^{hf/k_B T} - 1)^{-1}$, where $n_B$ is the magnon occupation density, $f$ is the magnon frequency, $h$ is the Planck constant, and $k_B$ is the Boltzmann constant. In the current study, our spin relaxometry experiments focus on YIG magnons at the lower spin transition ($m_s = 0 \leftrightarrow -1$) of $V_B^-$ electron spin resonance (ESR) frequency $f_-$ (*2, 35*). The observed $V_B^-$ spin relaxation is mainly driven by two independent mechanisms: spontaneous magnon emission and thermal magnon effects. The spontaneous emission describes flipping of a $V_B^-$ center from the $m_s = -1$ to $m_s = 0$ state that emits a magnon to the YIG spin bath at a rate of $\Gamma_0$ at zero temperature as illustrated in Fig. 2D (*39*). Using the fluctuation-dissipation theorem, $\Gamma_0$ can be expressed as (see Section S5 for details) (*35*):

$$\Gamma_0 = \frac{1}{t_{hBN}} \iint_{t_{\mathrm{Au}}}^{t_{\mathrm{Au}}+t_{\mathrm{hBN}}} D(f_-, \boldsymbol{k}) F(\boldsymbol{k}, t) d\boldsymbol{k} d\boldsymbol{t} \quad (1)$$

where $D$ is the magnon spectral density, $t$ is the distance between $V_B^-$ centers and the top surface of the YIG film, $t_{hBN}$ and $t_{Au}$ are the thickness of hBN flake and Au stripline, $F(\boldsymbol{k}, t)$ is the transfer function describing magnon-generated magnetic fields at $V_B^-$ center sites. The thermal effect characterizes dipole coupling between $V_B^-$ centers and a YIG magnon bath at a finite temperature, featuring direct magnon absorption and emission processes (at the $V_B^-$ spin energy) that induce $m_s = 0 \leftrightarrow -1$ spin relaxation (*35*). Such magnon dynamics can be understood by invoking the Holstein-Primakoff transformation involving magnon creation or annihilation when the $V_B^-$ frequency is above the YIG magnon bandgap energy. At a finite temperature, the thermally occupied magnon bath produces both absorption and stimulated-emission processes. For a single $V_B^-$ center, the downward transition $m_s = -1 \rightarrow 0$ occurs at rate $\Gamma_0(1 + n_B)$, where the factor 1 represents spontaneous magnon emission and $n_B$ represents stimulated emission. The upward transition $m_s = 0 \rightarrow -1$ occurs at rate $\Gamma_0 n_B$, corresponding to thermal magnon absorption. Such asymmetric spin relaxation can be experimentally revealed by $V_B^-$ PL measurements under different initialization and readout states (see Section S4 for details). The measured $V_B^-$ population relaxation rate is therefore the sum of the upward and downward rates:

$$\Gamma = \Gamma_0(1 + 2n_B) = \Gamma_0 \left(1 + \frac{2}{e^{hf_-/k_B T} - 1}\right) \quad (2)$$

Our experimental results are in satisfactory agreement with the theoretical predictions. In the near zero-temperature limit where the thermal energy is below the $V_B^-$ ESR frequency (~150 mK), the YIG magnon occupation density $n_B < 1$ and the magnon bath is only weakly populated, thus, the spontaneous emission effect is more pronounced than or dominates over the thermal effect (*2, 39*). We find that the measured $V_B^-$ spin relaxation rate Γ decreases as the temperature reduces and ultimately saturates to the value of spontaneous magnon emission rate $\Gamma_0$ when $k_B T \ll hf_-$ (Fig. 2C). In our quantum sensing experiments, we have prepared multiple samples with different thicknesses $t_{Au}$ of the on-chip Au stripline to control spatial separation between the hBN nanoflake and YIG sample. For hBN/YIG samples with $t_{Au}$ = 60-, 110-, and 160 nm, Γ qualitatively increases with the decreasing $t_{\mathrm{Au}}$, showing that a reduced $V_{\mathrm{B}}^{-}$-to-magnon distance will enhance the spontaneous magnon emission rate $\Gamma_0$, which is fundamentally governed by the $V_{\mathrm{B}}^{-}$-magnon dipole coupling (see Section S4 for details). The excellent agreement between experimentally obtained $\Gamma_0$ and the theoretical model across varying $t_{\mathrm{Au}}$, as summarized in Fig. 2E, corroborates observation of spontaneous magnon emission from $V_{\mathrm{B}}^{-}$ centers in the near-zero temperature limit.

In the "high-temperature" regime where $k_{\mathrm{B}}T \gg hf_{-}$, the measured spin relaxation is dominated by fluctuating magnetic fields generated from YIG thermal magnons. The Bose-Einstein distribution function is approximated to $n_{\mathrm{B}} \approx \frac{k_{\mathrm{B}}T}{hf_{-}}$, which is also referred to as the Rayleigh-Jeans distribution, and the overall YIG thermal magnon density is proportional to measurement temperature. Figure 2F plots $\Gamma$ as a function of temperature up to ~3.5 K. One can see that the $V_{\mathrm{B}}^{-}$ spin relaxation rate exhibits a linear temperature dependence when above ~500 mK for the three hBN/YIG samples with different values of $t_{\mathrm{Au}}$, highlighting the dominant thermal magnon effect when $k_{\mathrm{B}}T \gg hf_{-}$.

## Magnetic field control of spontaneous magnon emission of $V_{\mathrm{B}}^{-}$ centers

Since the $V_{\mathrm{B}}^{-}$-magnon dipole coupling serves as a driving force to the spontaneous magnon emission process, the measured $V_{\mathrm{B}}^{-}$ spin relaxation rates naturally inherit characteristics of the spin-wave dispersion of the magnetic material. Next, we tune the $V_{\mathrm{B}}^{-}$ spin transition frequency to show that the spontaneous magnon emission rate $\Gamma_0$ can be effectively controlled by the wavevector of the YIG magnon bath. Figure 3A plots a normalized spin wave transmission map of the 110-nm-thick YIG sample measured (see Section S6 for details). Note that the direction of the external magnetic field $B_{\mathrm{ext}}$ is 14° tilted away from the out-of-plane axis, which is the same as that in our $V_{\mathrm{B}}^{-}$ spin relaxometry measurements. At the ferromagnetic resonance (FMR) condition, a significant amount of microwave power is absorbed by the YIG magnetization, leading to a reduced microwave transmission efficiency characterized by $S_{12}$ (*40, 42, 43*). The FMR frequency $f_{\mathrm{FMR}}$ defining the minimum YIG magnon energy increases with $B_{\mathrm{ext}}$, and the energy difference between $f_{-}$ and $f_{\mathrm{FMR}}$ continuously decreases with increasing $B_{\mathrm{ext}}$ in the magnetic field range surveyed. Thus, it is expected that a YIG magnon spontaneously emitted by the $m_s = -1 \rightarrow 0$ spin transition of a $V_{\mathrm{B}}^{-}$ center tends to possess a larger wavevector in the low magnetic field regime and vice versa as shown by the YIG magnon wavevector $k$ map in Fig. 3B (see Section S6 for details). Figure 3C presents variations of $\Gamma_0$ as a function of the external magnetic field $B_{\mathrm{ext}}$. The measured $\Gamma_0$ shows a monotonical decrease with $B_{\mathrm{ext}}$ for the three hBN/YIG samples surveyed and nicely follows the theoretical prediction (Eq. 1), demonstrating field control of spontaneous magnon emission rate of $V_{\mathrm{B}}^{-}$ centers. A slower spin relaxation rate observed in the higher magnetic field regime is attributed to the $k$ dependent transfer function $F(\boldsymbol{k}, t)$ that enters in the calculation of $V_{\mathrm{B}}^{-}$ spin relaxation rate (*35*). In the limit of $k < 1/t$, a high-wavevector YIG magnon is energetically more favorable to be emitted from a $V_{\mathrm{B}}^{-}$ center, thus, a higher magnon energy ($f_{-}$) at a lower magnetic field will give rise to an enhanced magnon emission rate (see Section S5 for details). By setting the $V_{\mathrm{B}}^{-}$ ESR frequency to an arbitrary value in Eq. 1, we can further obtain a spontaneous magnon emission rate map as presented in Fig. 3D.

## Magnon-mediated correlations in $V_{\mathrm{B}}^{-}$ spin ensembles

The spontaneous magnon emission observed at millikelvin temperatures opens a promising route for engineering magnon-mediated correlations in $V_{\mathrm{B}}^{-}$ spin ensembles. When $V_{\mathrm{B}}^{-}$ defects are coupled through a magnon bath, spin relaxation of the ensemble contains two major contributions:

(i) the usual independent single-defect decay, and (ii) interference terms that correlate magnon emission (or absorption) events from different spin defects. In the dilute regime, the relative phase accumulated between defects $i$ and $j$, set by $e^{i\boldsymbol{k}\cdot(\boldsymbol{r}_i-\boldsymbol{r}_j)}$, varies rapidly from pair to pair, thereby suppressing inter-defect coherence; consequently, the ensemble relaxes essentially as a collection of independent defects (*2*). Here, $\boldsymbol{r}_i - \boldsymbol{r}_j$ is the separation vector between $V_{\mathrm{B}}^-$ defects. By contrast, in dense $V_{\mathrm{B}}^-$ ensembles with reduced inter-defect spacing, the magnon-mediated interference contributions can add constructively (*2*), potentially resulting in collective spin dynamics where spin relaxation ceases to be treated as independent depolarization of $V_{\mathrm{B}}^-$ centers and instead evolves to a correlated state with an enhanced relaxation rate as illustrated in Fig. 4A.

Figure 4B summarizes the estimated $V_{\mathrm{B}}^-$ spin density $\rho$ of four hBN flakes with different neutron irradiation doses (see Section S1 for details). Figure 4C presents four representative $V_{\mathrm{B}}^-$ spin relaxation spectra recorded at 90 mK. The $V_{\mathrm{B}}^-$ spin ensembles follow nearly identical relaxation behavior for $\rho_1 = 1.7 \times 10^{17}$ cm$^{-3}$ and $\rho_2 = 3.1 \times 10^{17}$ cm$^{-3}$. As the spin defect density increases to $\rho_3 = 6.8 \times 10^{17}$ cm$^{-3}$ and $\rho_4 = 1.7 \times 10^{18}$ cm$^{-3}$, the PL spectra of $V_{\mathrm{B}}^-$ spin ensembles exhibit a faster decay, reflecting an enhanced relaxation rate. By fitting our results to an exponential decay function, magnon-driven $V_{\mathrm{B}}^-$ spin relaxation rate $\Gamma$ can be obtained as summarized in Fig. 4D (see Section S4 for details). We have developed a correlated spin defect model to account for spin defect-density-dependent $V_{\mathrm{B}}^-$ relaxation. Correlated $V_{\mathrm{B}}^-$ spin relaxation rate can be approximately described by the following expression (see Sections S7 and S8 for details):

$$\Gamma = \sqrt{(A+B)^2 - 4AC} \tag{3}$$

where $A = \eta\Gamma_0\lambda^2 t_{\mathrm{hBN}}\rho$ represents the correlated magnon emission contribution, $B = \Gamma_0(1 + 2n_{\mathrm{B}})$ is the uncorrelated spin relaxation rate considering both spontaneous emission and thermally driven transitions, and $C = \Gamma_0 n_{\mathrm{B}}$ characterizes the contribution from thermal magnon absorption. Note that $\lambda$ is the wavelength of the magnons resonant with the spin-defect transition frequency $f_-$, and $\eta$ is a phenomenological parameter which rescales spin defect number and defines an effective $V_{\mathrm{B}}^-$ density in our model. When $\eta \to 0$, our model reduces to the independent spin dynamics and $\Gamma \to \Gamma_0(1 + 2n_{\mathrm{B}})$. Non-idealities such as local dephasing, inhomogeneous broadening, nonuniform coupling strengths, and disorder will suppress the correlation and further reduce from its ideal value (see Section S7 for details). Substituting relevant material parameters into the above equation, theoretical calculations of $\Gamma$ are presented in Fig. 4D.

Overall, our experimental results are in agreement with the model. For $\rho_1 = 1.7 \times 10^{17}$ cm$^{-3}$ and $\rho_2 = 3.1 \times 10^{17}$ cm$^{-3}$, the measured $\Gamma$ basically follows the theoretical prediction of $\Gamma_0(1 + 2n_{\mathrm{B}})$, indicating a largely uncorrelated $V_{\mathrm{B}}^-$ spin relaxation behavior in the dilute regime. In the high defect density regime, distinct $V_{\mathrm{B}}^-$ centers couple to resonant modes of the YIG magnon bath, enabling the bath to mediate nonlocal decay channels between different spin defects. Under these conditions, reservoir-induced cross couplings give rise to correlated spin relaxation dynamics, manifested as density-dependent deviations from the independent-defect decay (*2*). For $\rho_3 = 6.8 \times 10^{17}$ cm$^{-3}$, measured $V_{\mathrm{B}}^-$ spin relaxation rates are enhanced from the uncorrelated case due to the pronounced correlated magnon emission effect. Intuitively, a higher $\rho$ reduces the typical defect-defect separation, so a single magnon mode can couple a larger number of defects within its spatial reach. When spin defect density increases to $\rho_4 = 1.7 \times 10^{18}$ cm$^{-3}$, the magnon-mediated correlation effect in $V_{\mathrm{B}}^-$ ensembles becomes more pronounced, as evidenced by the further increase in $\Gamma$. When normalized by the uncorrelated relaxation value, $\frac{\Gamma}{\Gamma_0(1+2n_B)}$ features significant enhancement in the sub-Kelvin regime as shown in Fig. 4E. As expected, this enhancement is suppressed with increasing temperature. When $T > 1$ K, signatures of magnon-mediated correlation fully disappear, which is attributed to the dominant thermal magnon effect that overshadows correlated magnon emission of $V_{\mathrm{B}}^{-}$ centers (*2*), resulting in absence of density dependent effects and convergence of decay rates across different samples.

## Discussion

In summary, we have demonstrated millikelvin quantum sensing using boron-vacancy spin defects in 2D hBN nanoflakes. When lowering from $^4$He cryogenic temperatures to the millikelvin regime, we observe orders of magnitude decrease of the intrinsic $V_{\mathrm{B}}^{-}$ spin relaxation rate due to suppression of phonon scattering-induced spin depolarization. Taking advantage of the extended $V_{\mathrm{B}}^{-}$ spin relaxation times, we introduce a shared YIG magnon bath to investigate dipole coupling between $V_{\mathrm{B}}^{-}$ centers and a ferromagnetic spin bath, presenting clear evidence of spontaneous magnon emission from $V_{\mathrm{B}}^{-}$ centers in near zero-temperature limit. By varying the spin defect density, we further explore experimental signatures of magnon-mediated correlated quantum spin relaxation dynamics of $V_{\mathrm{B}}^{-}$ ensembles. These results, reproduced well by our theoretical modeling, bring new insights into understanding collective spin dynamics, magnon-qubit coupling and many-body physics of spin defects in the quantum regime (*2*). Our work also highlights the potential of magnons as mediating quanta capable of establishing long-range, tunable interaction between individual color centers and driving cooperative quantum effects such as super- and subradiant dynamics in strongly correlated spin defect ensembles (*1, 4, 5, 31, 44-47*). The hybrid systems consisting of magnons and hBN spin defects with tailored quantum mechanical properties also open new avenues for developing cutting-edge technological innovations from sensing, metrology, and potentially to computing at the forefront of quantum revolution (*11, 32, 48-50*).

**Acknowledgements**: The authors are grateful to Yaroslav Tserkovnyak for insightful discussions.

**Funding:** The quantum sensing measurements are supported by the Office of Naval Research (ONR) under grant No. N00014-23-1-2146. Development of millikelvin quantum microscopy is based upon work supported by the Air Force Office of Scientific Research under award number FA9550-25-1-0082. Device fabrications are supported by U.S. National Science Foundation under award No. DMR-2437294, No. ECCS-2445826 and No. ECCS-2525800. B. F. and J. M. P. N. acknowledge support from DOE under Award No. DE-SC0024090. J. H. E and T. P. thank the National Science Foundation for supporting the h$^{10}$B$^{15}$N crystal growth under award No. 2413808. Neutron irradiation of the h$^{10}$B$^{15}$N is supported by the U.S. Department of Energy, Office of Nuclear Energy, under DOE Idaho Operations Office Contract DE-AC07-05ID14517 as part of Nuclear Science User Facilities award No. 24-4911. S. R. and Z. J. thank the NASA Solar System Exploration Research Virtual Institute (SSERVI), under cooperative agreement number NNH22ZDA020C (CLEVER, Grant number: 80NSSC23M022) for synthesis of $^{10}$B-enriched monoisotopic hBN crystals.

**Author contributions:** L.-J. Z. performed quantum sensing measurements and analyzed the data with S. L. J. M. P. N. and B. F. developed the theory on magnon-mediated correlated relaxation dynamics of spin ensembles. Y. Z., Z. X. and J. Z. contributed to device fabrication processes. T. P. and J. H. E. provided isotopically purified h$^{10}$B$^{15}$N crystals with different neutron irradiation doses. S. R. and Z. J. provided monoisotopic h$^{10}$BN crystals. H. W. contributed to discussion and data interpretation. C. R. D. supervised this project.

**Competing interests:** The authors declare no competing interests.

**Data, code, and materials availability**: All data and code needed to evaluate the conclusions in the paper are present in the paper and/or the Supplementary Materials. Request for the isotopically purified h$^{10}$B$^{15}$N crystals used in this study are available from Kansas State University pending a materials transfer agreement. Requests can be made to edgarjh@k-state.edu.

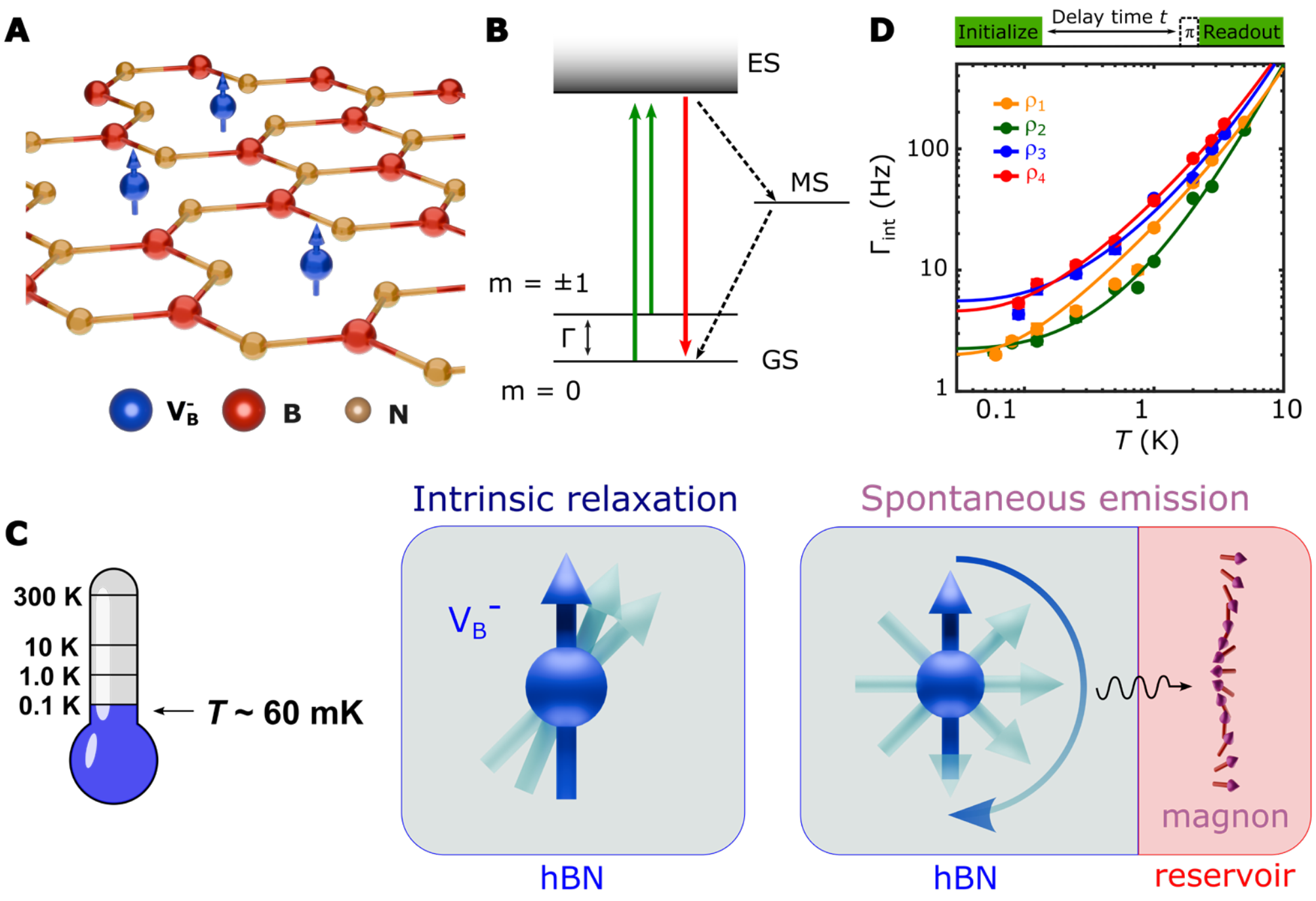

**Figure 1. Quantum spin relaxation of boron-vacancy centers in near zero-temperature limit.** (**A**) $V_B^-$ spin defects (blue) formed in the hBN crystal structure with alternating boron (red) and nitrogen (brown) atoms. (**B**) Energy level diagram of a $V_B^-$ spin defect and the optical excitation (green arrow), radiative recombination (red arrow), and nonradiative decay (black dashed arrows) between the ground state (GS), excited state (ES), and metastable state (MS). (**C**) Schematics of intrinsic $V_B^-$ spin relaxation dynamics and spontaneous magnon emission to a proximal spin reservoir in the near-zero temperature limit. (**D**) Temperature dependence of intrinsic $V_B^-$ spin relaxation rate $\Gamma_{int}$ measured down to ~60 mK for $V_B^-$ spin ensembles with 4 different spin defect densities ($\rho_1$, $\rho_2$, $\rho_3$, and $\rho_4$). Experimental results (dots) are fitted to the theoretical model (curves). Optical and microwave sequence for spin relaxometry measurements is depicted on top.

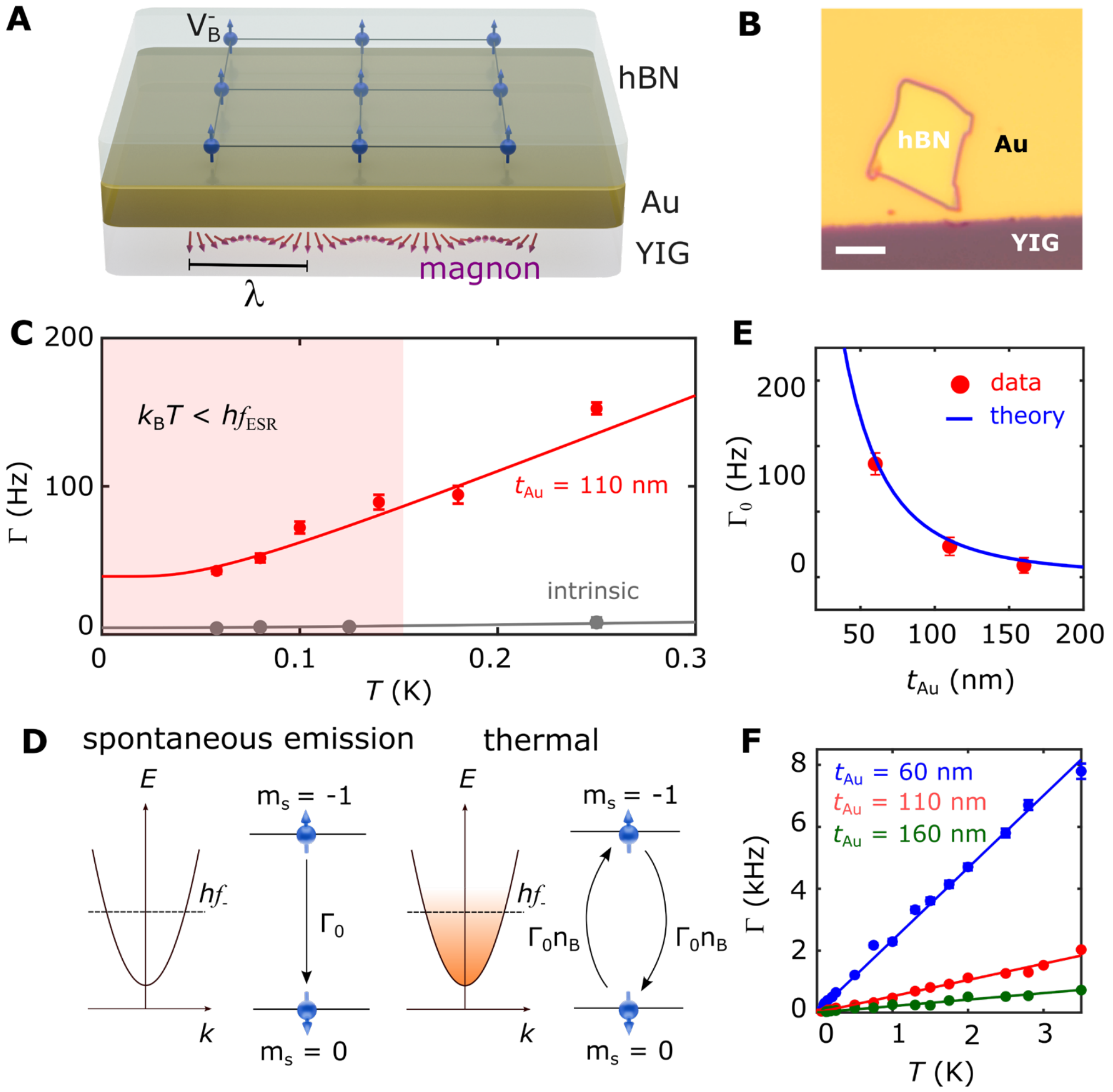

**Figure 2. Observation of spontaneous magnon emission from $V_B^-$ centers.** (**A**) Schematic view of our measurement platform where a hBN nanoflake containing $V_B^-$ spin defects is transferred on top of an on-chip Au microwave stripline patterned on a 110-nm-thick YIG film. The spatial distribution of $V_B^-$ spin defects is plotted in an illustrative way and not to scale. (**B**) Optical microscopy image of a prepared hBN/YIG device. The scale bar is 5 μm. (**C**) Temperature dependent $V_B^-$ spin relaxation rate Γ measured on a hBN/YIG sample in the sub-Kelvin regime. The estimated $V_B^-$ spin defect density is $\rho_2 = 3.1 \times 10^{17}$ cm$^{-3}$ and the thickness of Au stripline is $t_{Au} = 110$ nm. The intrinsic $V_B^-$ spin relaxation rate (gray color data points) is also presented for reference. (**D**) Schematic illustration of spontaneous magnon emission and thermal magnon effects on $V_B^-$ spin relaxation in zero- and finite temperature regime. YIG magnon dispersion and band structure are plotted, and energy dependent magnon density is highlighted by the fading colors. (**E**) Spontaneous magnon emission-induced $V_B^-$ spin relaxation rate $\Gamma_0$ as a function of the Au stripline thickness $t_{Au}$. (**F**) Magnon-$V_B^-$ dipole coupling-driven spin relaxation rate Γ measured up to ~3.5 K for three hBN/YIG samples with the same $V_B^-$ spin defect density ($\rho_2 = 3.1 \times 10^{17}$ cm$^{-3}$) and different values of $t_{Au}$. Points in (C) and (F) denote experimental results and solid lines (curves) represent theoretical calculations.

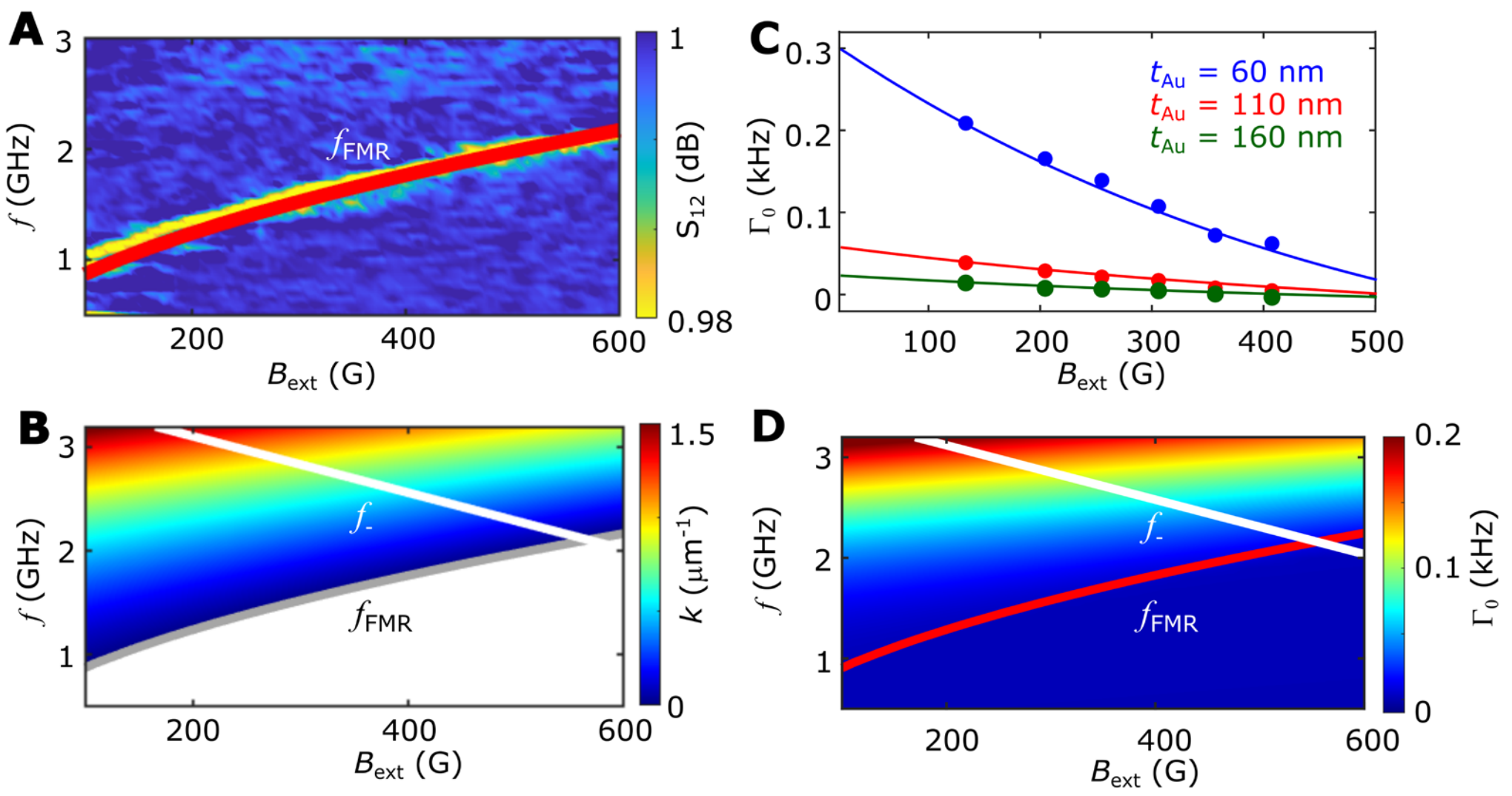

**Figure 3. Magnetic field control of spontaneous magnon emission rate of $V_B^-$ centers.** (**A**) FMR spin wave transmission map of a 110-nm-thick YIG film measured at 500 mK. The red curve highlights the YIG FMR frequency. (**B**) Theoretically calculated YIG magnon wavevector map with white and gray curves representing the $V_B^-$ ESR frequency $f_-$ ($m_s = 0 \leftrightarrow -1$) and the YIG FMR frequency $f_{FMR}$. (**C**) Spontaneous magnon emission rate $\Gamma_0$ measured as a function of the external magnetic field $B_{ext}$. The blue, red, and green data points correspond to three hBN/YIG samples with 60-, 110-, and 160-nm-thick on-chip Au striplines. The $V_B^-$ spin defect density is estimated to be $\rho_2 = 3.1 \times 10^{17}$ cm$^{-3}$ for the three hBN samples surveyed. Points are experimental results and solid lines present theoretical calculations. $\Gamma_0$ values are extracted from temperature dependent $V_B^-$ spin relaxometry measurement down to sub-100 mK. (**D**) Theoretically calculated magnon emission rate $\Gamma_0$ map with white and red lines denoting the $V_B^-$ ESR energy $f_-$ ($m_s = 0 \leftrightarrow -1$) and the YIG FMR frequency $f_{FMR}$, respectively.

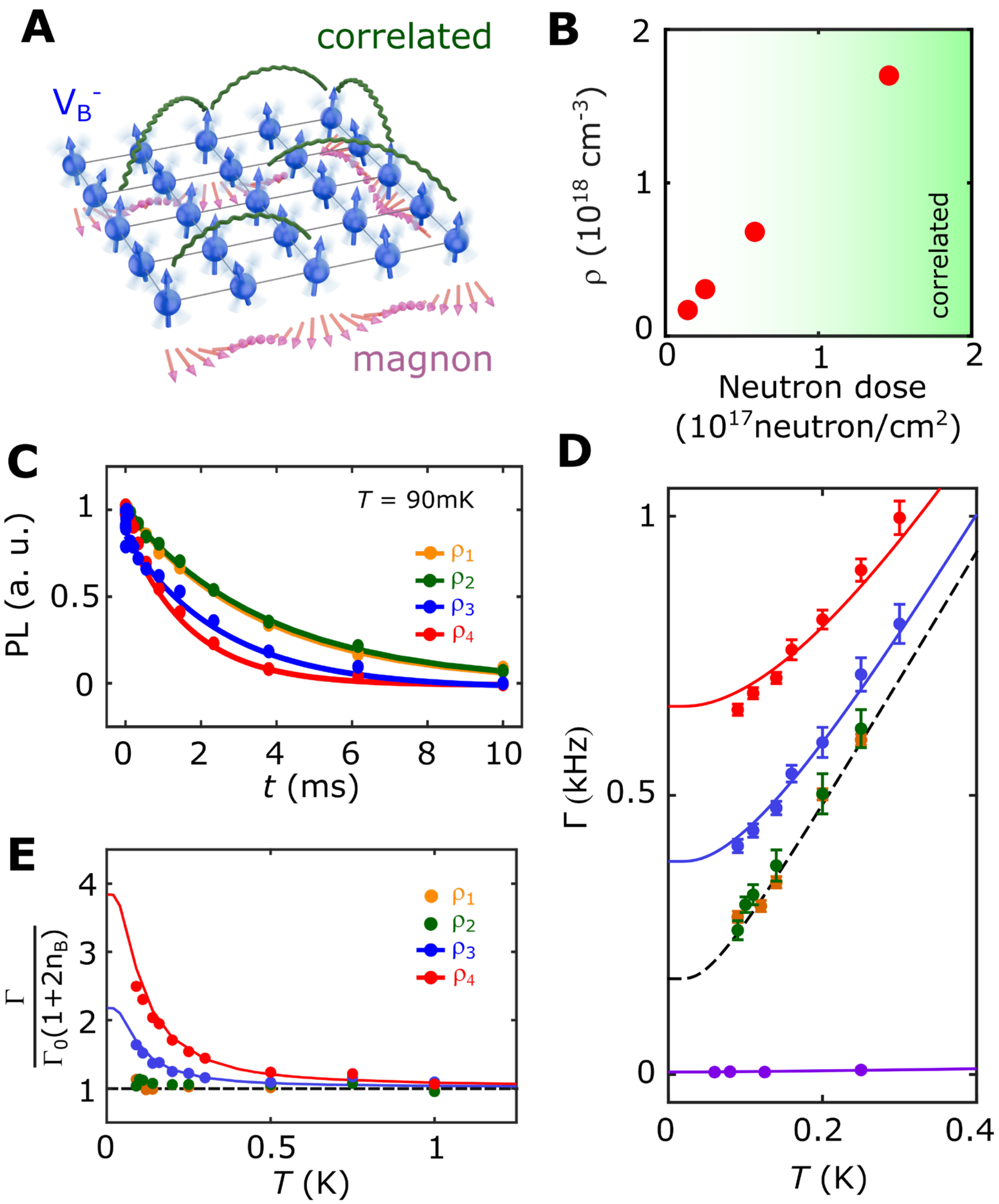

**Figure 4. Magnon-mediated correlated $V_B^-$ spin relaxation.** (**A**) Schematics of correlated relaxation of $V_B^-$ spin ensembles when interfaced with a shared magnon bath. (**B**) $V_B^-$ spin defect density $\rho$ estimated as a function of the neutron irradiation dose. The fading green color indicates gradual transition from uncorrelated to correlated spin relaxation regime with increasing $\rho$. (**C**) Spin relaxation spectra of $V_B^-$ ensembles with four different spin defect densities ($\rho_1$, $\rho_2$, $\rho_3$, and $\rho_4$) measured at 90 mK. Experimental results (dots) are fitted to exponential decay (lines). The thickness of the Au stripline is 60 nm for the four hBN/YIG samples surveyed. (**D**) Spin relaxation rate $\Gamma$ of $V_B^-$ ensembles measured as a function of temperature in the sub-Kelvin regime. Results on four hBN samples with different $V_B^-$ spin defect densities are presented. Black dashed lines

represent theoretically calculated uncorrelated $V_B^-$ spin relaxation rate $\Gamma_0(1+2n_B)$, and intrinsic spin relaxation rate $\Gamma_{int}$ is also presented for reference. Experimental data sets of $\rho_3$ and $\rho_4$ are fitted to theoretical calculations (blue and red solid lines). (**E**) Magnon-driven $V_B^-$ spin relaxation rate Γ normalized by the uncorrelated $V_B^-$ spin relaxation rate in temperature range up to 1.25 K. Red and blue lines denote theoretical calculations. Black dashed lines represent the value of 1 for visual clarity.